# Social network analysis of *Staphylococcus aureus* carriage in a general youth population

Dina B. Stensen, MD *[1,2], Rafael A. Nozal Cañadas, MSc *[3], Lars Småbrekke, PhD [4], Karina Olsen, PhD [5], Christopher Sivert Nielsen, PhD [6,7], Kristian Svendsen, PhD [4], Anne Merethe Hanssen, PhD [8], Johanna UE Sollid, PhD [8], Gunnar Skov Simonsen, PhD [5,8], Lars Ailo Bongo, PhD [3], Anne-Sofie Furberg, PhD [5,9]

[1]Department of Community Medicine, Faculty of Health Sciences, UiT The Arctic University of Norway, Hansine Hansens veg 18, 9019 Tromsø, Norway

[2]Division of Internal Medicine, University Hospital of North Norway, Sykehusvegen 38, 9019 Tromsø, Norway

[3]Department of Computer Science, UiT The Arctic University of Norway, Hansine Hansens veg 54, 9019 Tromsø, Norway.

[4]Department of Pharmacy, Faculty of Health Sciences, UiT The Arctic University of Norway, Hansine Hansens veg 18, 9019 Tromsø, Norway

[5]Department of Microbiology and Infection Control, Division of Internal Medicine, University Hospital of North Norway, Sykehusvegen 38, 9019 Tromsø, Norway

[6]Department of Chronic Diseases and Ageing, Norwegian Institute of Public Health, Marcus Thranes gate 6, 0473 Oslo, Norway

[7]Department of Pain Management and Research, Division of Emergencies and Critical Care, Oslo University Hospital, Postboks 4956 Nydalen, 0424 Oslo, Norway

[8]Department of Medical Biology, Faculty of Health Sciences, UiT The Arctic University of Norway, Hansine Hansens veg 18, 9019 Tromsø, Norway

[9]Faculty of Health and Social Sciences, Molde University College, Britvegen 2, 6410 Molde, Norway.

 *These authors have contributed equally to the development of the manuscript

Corresponding author: Dina B. Stensen.

Institution: Department of Community Medicine, Faculty of Health Sciences, UiT The Arctic University of Norway.

Postal address: Hansine Hansens veg 18, 9019 Tromsø, Norway. E-mail: dina.b.stensen@uit.no

# Abstract

Background

*Staphylococcus aureus* nasal carriage increases risk of infection and has been associated with lifestyle behavior and biological host characteristics. We used social network analysis to evaluate whether contacts have the same *S. aureus* genotype indicating direct social transmission, or whether contagiousness is an indirect effect of contacts sharing the same lifestyle or characteristics.

Methods

The Fit Futures 1 study collected interview data on social contact among 1038 first level students in the same high school district in Norway. *S. aureus* persistent carriage was determined from two nasal swab cultures and *S. aureus* genotype from *spa*-typing of a positive throat swab culture. Bootstrap, t-tests, logistic regression, and autocorrelation were used to evaluate social network influence on host risk factors and *S. aureus* carriage.

Findings
Both *S. aureus* persistent carriage and *spa*-type were transmitted in the social network ($p<0.001$). The probability of carriage increased by 3·7% and 5·0% for each additional *S. aureus* positive friend, in univariable regression and multivariable autocorrelation analysis respectively. Male sex was associated with a 15% lower risk of transmission compared to women, although the prevalence of carriage was higher for men (36% versus 24%). Medium physical activity, medium and high alcohol-use, and normal-weight students had higher number of contacts, and increased risk of transmission ($p<0.002$).

Interpretation

We demonstrate direct social transmission of *S. aureus* in a general youth population. Lifestyle factors are associated with risk of transmission suggesting indirect social group effects on *S. aureus* carriage from friends having more similar environmental exposures. The male predominance in carriage is determined by sex-specific predisposing host characteristics as *S. aureus* social transmission is less frequent than in females. Better understanding of how social interactions influence *S. aureus* carriage dynamics in the population is important for developing new preventive measures.

Funding

The Northern Norway regional Health Authorities (grant number HNF1457-19).

# Introduction

Nasal carriage of *Staphylococcus aureus* (*S. aureus*) has a prevalence of 20-30% in the general adult population (1, 2) and 40-50 % in older children and adolescents (3), and is more common among men than among women (2). Carriers have an increased risk of autoinfection (4, 5). Prevention and eradication of nasal carriage may therefore reduce the *S. aureus* disease burden (5). Epidemiological studies have searched for modifiable risk factors for *S. aureus* nasal carriage as potential targets for interventions, including body mass index (BMI), serum glucose and vitamin D concentration, exogenous and endogenous hormone exposure, and smoking (2-4, 6-8). However, these studies did not adjust for social contact.

*S. aureus* is mainly directly transmitted through physical contact (9), but no other study has evaluated the direct social transmission of *S. aureus* carriage in a general population. In studies of transmissible pathogens, an extensive problem with identifying risk factors is the lack of adjustment for social contact. Biological host risk factors for *S. aureus* carriage may also be determinants of friendship, thereby producing an association by confounding. Predisposing lifestyle risk factors may be contagious with the consequence of researchers incorrectly assuming transmission of the pathogen. Prevention of *S. aureus* carriage is dependent on identifying key transmission pathways and causal risk factors to correctly evaluate targets for interventions.

Infectious diseases like tuberculosis, HIV infection and sexually transmitted diseases have been strongly connected with social networks (10-12). These studies demonstrate that the degree and type of contact between individuals play a significant role in disease incidence. One study showed that introduction of *S. aureus* into a social network of active drug users created a reservoir for the bacteria with linkages to the general population (13). A recent case-control study used network analysis to reveal transmission of methicillin-resistant *S. aureus* (MRSA) through a social network in healthcare (14). Social group effects also occur in humans as unrelated individuals living in the same household were found to have a more similar microbiota than relatives living in different households (15).

The aim of this study is therefore to estimate the extent to which *S. aureus* carriage follows friendship ties, and if the data support the concept of direct social transmission. We also want to identify host risk factors for *S. aureus* carriage and differentiate between the risk attributable to social contact between similar individuals compared to biological or lifestyle related risk per se.

# Methods

## Population and study design

The Fit Futures 1 study (FF1) is a youth health survey conducted from September 2010 to April 2011. The study invited all the 1117 first-year students registered at the eight high schools in the municipalities of Tromsø and Balsfjord, North-Norway, and reached 93% participation. Altogether, 508 female and 530 male students attended (16). Participants had a half-day visit at the Clinical Research Unit of the University Hospital of North Norway (UNN), including clinical examinations, microbiological samples, blood samples, a web-based questionnaire, and an interview. Trained research nurses did all procedures according to a standard protocol.

## Host risk factors

Height and weight were measured on an electronic scale with participants wearing light clothing and no footwear. BMI was calculated as weight (kg) divided by the squared height ($m^2$). The participants reported their sex, age, study program, tobacco use, alcohol use, and recreational physical activity through the web-based questionnaire.

The interview covered current hormonal contraceptive use. We categorized hormonal contraceptive use into combination contraceptives with high or low ethinylestradiol dosage and progestin-only contraceptives (3).

## Assessment of *S. aureus* carriage

The research nurses took a first set of nasal and throat swab samples at the hospital, and a second set at school after a mean interval of 17 days. Both vestibules nasi were sampled using the same NaCl (0·9%)-moistened sterile rayon-tipped swab, and both tonsillar regions were sampled with an additional swab. The swabs were immediately placed in transport medium (Amies Copan, Brescia, Italy) and stored at 4°C for a maximum of 3 days. All samples were analysed at the Department of Microbiology and Infection Control, UNN, both by direct culture (2) and enrichment culture (3) (Bacto Staphylococcus medium broth, Difco laboratories, Sparks, MD, USA), using blood agar for growth control (Oxoid, UK) and chromID-plates for *S. aureus* detection (SAID, bioMérieux, Marcy I'Etoile, France). Growth of any bacterial colonies on agar plates was registered as a valid sample. The dominating *S. aureus* colony type was frozen at - 70 °C in glycerol-containing liquid medium after confirmation by Staphaurex plus agglutination test (bioMérieux, Marcy I'Etoile, France).

We used *S. aureus* persistent nasal carriage as the main outcome variable in the present analysis as this has been the major phenotype of interest in infection control and epidemiological studies (17). We defined persistent carriage as having two *S. aureus* positive nasal cultures and created two different persistent carriage variables based on either direct culture or enrichment culture respectively (Supplementary Figure 1). Genotype data was available only for a subset of the nasal bacterial isolates. However, all *S. aureus* isolates from the first throat swab sample taken at the hospital were *spa*-typed as part of another study by our group. The frozen throat cultures were inoculated on blood agar (Oxoid) and incubated overnight at 37°C. Two or three colonies were transferred to 200 µl sterile $H_2O$ and vortexed. The isolates were *spa*-typed following the protocol described in Sangvik et al. (8).

Social network

We constructed the social network based on the interview question: "Which first level high school students have you had most contact with the last week? Name up to five students at your own school or other schools in Tromsø and Balsfjord." Reciprocity in the nomination was not mandatory. For each nomination, five "yes/no" questions assessed the type of contact: "Did you have physical contact?", "Have you been together at school?", "Have you been together at sports?", "Have you been together at home?", "Have you been together at other places?". This resulted in five social networks depending on setting: "physical contact", "school", "sport", "home, and "other" networks. Adding all the relationships together formed a sixth network that was called the "overall" network. To evaluate if the friends mentioned were representative for the participants´ social network, the following question was asked: "To what degree does this list of friends give an overview of your social network? Please indicate on a scale from zero (small degree) to ten (high degree)." We excluded 134 nominated friends that did not participate in FF1 from the analysis.

Statistical analysis

We used R version 3·6·3 and R Studio 1·3·1093 for the statistical analysis. To evaluate univariable associations between host risk factors and *S. aureus* persistent carriage we used t-test and chi-square test, with Yates's correction for 2x2 tables and Fisher's exact test, when applicable.

In the social network analysis, nodes refer to participants in the social network while edges refer to lines representing relationships between participants. To evaluate transmission of *S. aureus* through the social network, we analyzed edges between nodes using Exponential Random Graph Models (ERGM) or additive and multiplicative effects models. We analyzed patterns of connections (non-carriers connected to non-carriers, non-carriers connected to carriers, carriers connected to carriers) using Simulation Investigation for Empirical Network Analysis (SIENA), an autocorrelation model (18). In further analysis we used bootstrapping of simulated networks against the observed network, descriptive analysis, and logistic regression to evaluate the effect of host risk factors. The statistical background for our methods is described in the supplementary material.

## Ethics

Each participant in FF1 signed a declaration of consent. Participants younger than 16 years of age had to bring written consent from a parent or guardian. FF1 was approved by The Regional Committee of Medical and Health Research Ethics North Norway (reference 2009/1282) and the Norwegian Data Protection Authority. The present study was approved by REK North (reference 2011/1710).

## Role of the funding source

The funding sources were not involved in the study design, data collection, data analysis, data interpretation, writing of the paper, or any aspect pertinent to the study.

# Results

## Transmission of *S. aureus* carriage in a general population

In this general population with mean age = 16·4 years (SD = 1·24, range 15-28 years), the prevalence of *S. aureus* persistent carriage determined by direct culture was 30·3%, compared to 42·6% when using enrichment culture. Prevalence of persistent carriage was higher in male participants compared to female (Prevalence: 36·4% and 48·1%, men direct culture and enrichment culture; Prevalence: 24·0% and 36·8%, women direct culture and enrichment culture). We found no other significant differences between groups according to population characteristics (Supplementary Table 1).

We first evaluated the FF1 social network structure based on all relationships between students in the five subnetworks (Supplementary Figure 2) and information about relationships and persistent carriage status of nodes in the "overall" network diagram (Supplementary Figure 3). Because the population was recruited from two neighboring municipalities, there are two distinct clusters of students. In a truly random network, it is expected that the total number of edges between high school clusters would approximate the number of edges within the same cluster. In this population, the number of edges inside a high school cluster was higher than outside of the cluster demonstrating high school as a strong driver of friendship (homophily of 87·8) (Figure 1). Likewise, participants tended to bond with similar students with respect to sex and lifestyle risk factors (Supplementary Table 2 and Supplementary Figure 4).

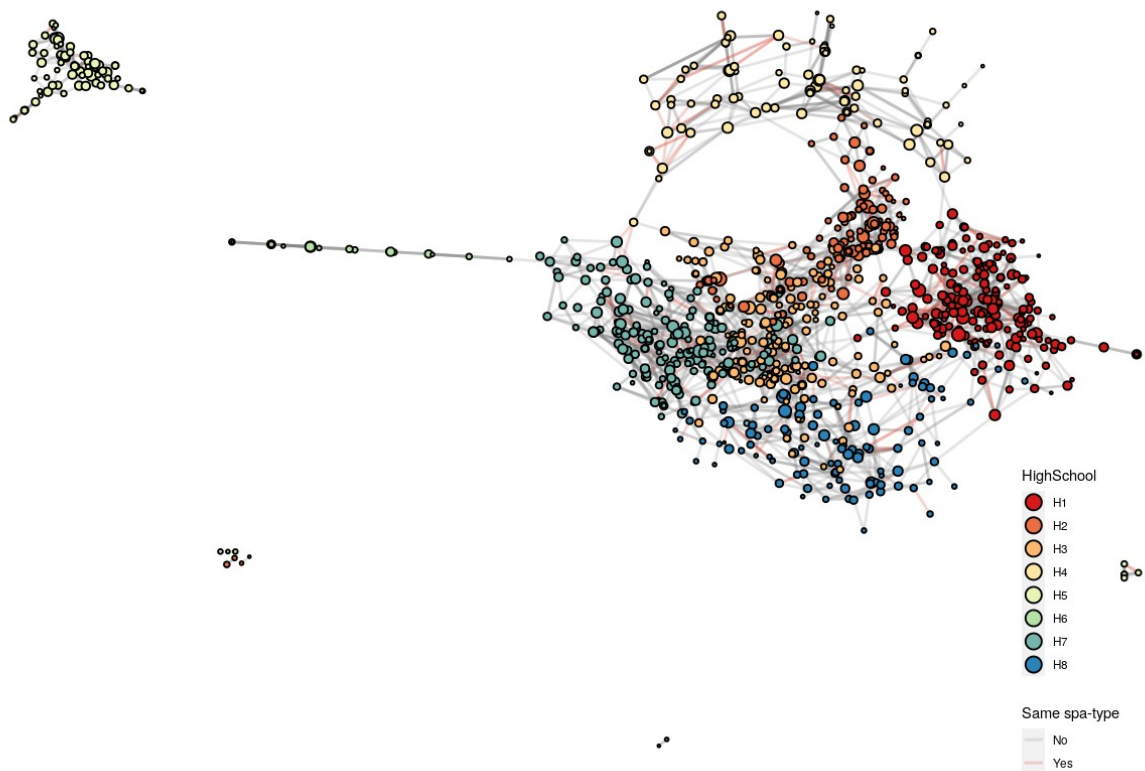

**Figure 1 The overall social network within and between high schools, with a multidimensional scaling layout. The Fit Futures 1 study.** Edges (lines) connecting nodes (students) with the same *S. aureus spa*-type in throat culture are drawn in red. Edges connecting nodes with different *spa*-type are drawn in grey. High school ID (H1-H8) represents the eight high schools included in Fit Futures 1. H5 represents students at the high school in Balsfjord municipality (isolated cluster, upper left). All other high schools (H1-H4, H6-H8) are in Tromsø municipality. Only students with *Staphylococcus aureus* isolated by direct or enrichment culture from the first throat swab sample are shown (N = 746). Unconnected students are not included (N = 21).

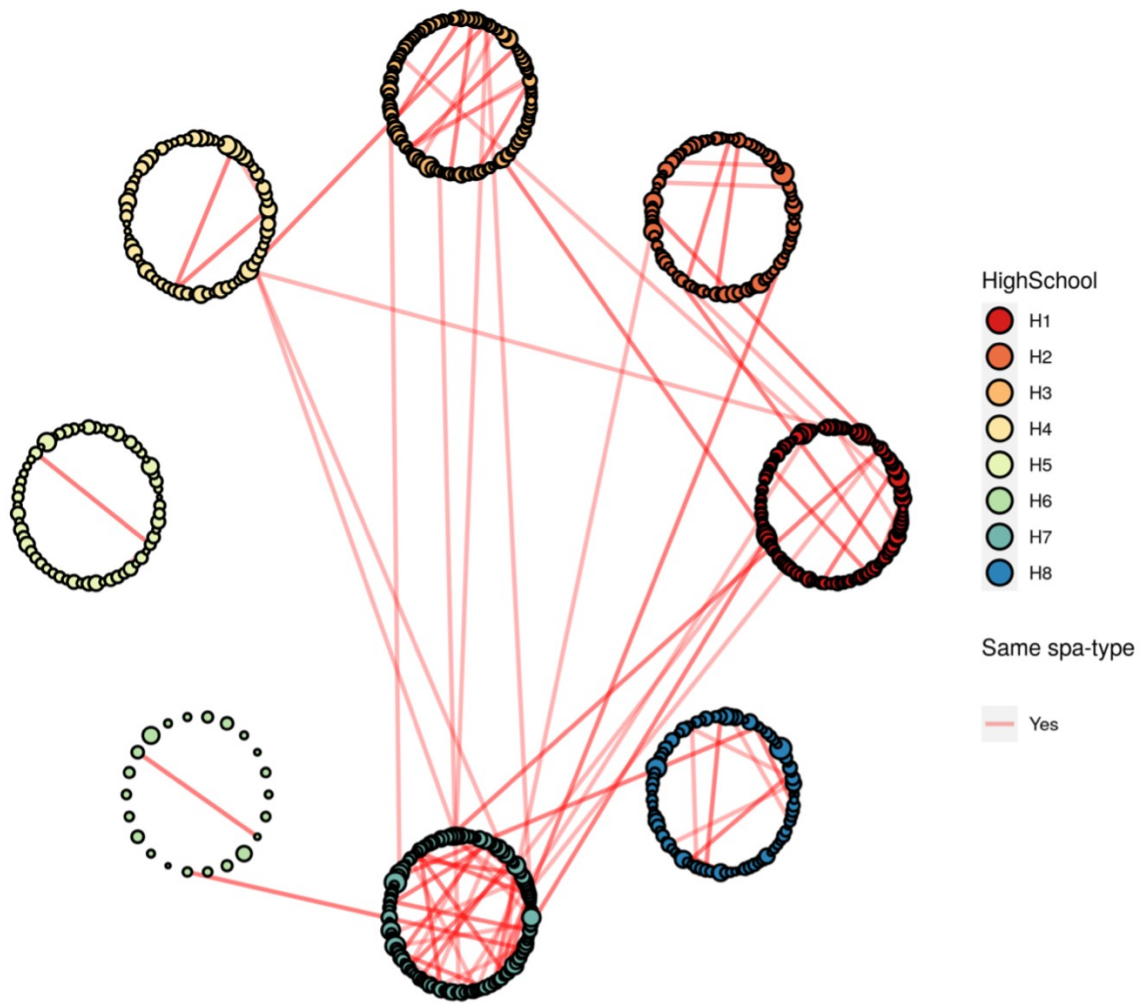

**Figure 2 The overall social network with students grouped by high school**. **The Fit Futures 1 study**. Edges (lines) connecting nodes (students) with the same *S. aureus spa*-type in throat culture are drawn in red. Edges connecting nodes with different *spa*-types are hidden. High school ID (H1-H8) represents the eight high schools included in the Tromsø Study Fit Futures 1. H5 represents students at the high school in Balsfjord municipality. All other high schools (H1-H4, H6-H8) are in Tromsø municipality. Only students with *Staphylococcus aureus* isolated by direct or enrichment culture from the first throat swab sample are shown (N = 746). Unconnected students are not included (N = 21).

The role of host risk factors in *S. aureus* transmission

In the logistic regression analysis, female participants had the highest risk of being exposed to *S. aureus* through their social interaction (Table 2). Men had a relative risk of transmission of 0·85 compared to women (95% confidence interval (CI)=0·805-0·884). Also, students using alcohol two times or more per month had a higher risk of transmission of *S. aureus* compared to students using alcohol once per month or less (p=0·035; direct culture). There was a higher probability of transmission among participants doing medium level physical activity (p=0.008) compared with the light physical activity group.

Female students had an average number of friends of 3·81, which was significantly higher than the average 3·46 friends among male students (p = 0·008) (Supplementary Table 5). Students consuming alcohol more than two times a month had higher number of friends compared to those consuming less or no alcohol (p<0·001).

We demonstrate that students who had a higher number of friends being persistent carriers were more likely to be persistent carriers themselves. This was significant for persistent carriage defined by both direct culture (p=0·002) and enrichment culture (p<0·001) (Supplementary Table 6). The probability of being a carrier increased by 3·7% (95% CI=3·52-3·94; univariable logistic regression, result not presented in table) when increasing your friend circle by one *S. aureus* positive friend defined by direct culture, as also illustrated in Figure 3 and Supplementary Table 7. Similarly, the probability increased by 3·4% (95% CI=3·33-3·45) for persistent carriage defined by enrichment culture (result not presented in table).

An adapted linear autocorrelation analysis gave comparable results to the logistic regression analysis (Table 3). Adjusting for host risk factors, the probability of persistent carriage increased by 4·8% (p<0·001) for each additional *S. aureus* positive friend defined by direct culture. A similar increase of 6% (p<0·001) was shown for the enrichment culture. The autocorrelation model also assessed which risk factors made the participants´ friends significantly more contagious. For direct culture there was an association for sex, BMI and physical activity (p=0·001-0·008), and for enrichment culture; study program, BMI and physical activity (p< 0·001 for all). Because of the assumptions of the autocorrelation model, beta estimates for individual host factors cannot be interpreted and sex-specific host factors (hormonal contraceptives) could not be included in the model. Females tended to have more relationships than males, which is also true for participants with normal BMI and participants with both medium and hard level physical activity (Supplementary Table 5).

**Table 1 Associations between host risk factors and transmission of *Staphylococcus aureus* persistent nasal carriage in the overall social network.** Results from two different regression analyses producing "P-value" for the social effect of each characteristic and Relative risk for the comparison of risk of transmission between groups. Persistent nasal carriage determined by both direct culture and enrichment culture. The Fit Futures 1 study (N = 1038).

| Risk factor (categories) | | Direct culture | | | Enrichment culture | | |
|---|---|---|---|---|---|---|---|
| | | P-value | Relative Risk | 95% CI | P-Value | Relative Risk | 95% CI |
| Sex | Female | **0.0002** | 1 | | **0.027** | 1 | |
| | Male | 0.999 | **0.845** | **0.805 - 0.884** | 0.843 | **0.937** | **0.900 - 0.974** |
| Study program | Vocational | 0.548 | 1 | | 0.353 | 1 | |
| | General | 0.510 | 1.002 | 0.950 - 1.054 | 0.410 | 0.996 | 0.954 - 1.038 |
| | Sport | 0.403 | 1.009 | 0.960 - 1.059 | 0.811 | 0.974 | 0.935 - 1.013 |
| BMI[a] | Underweight | 0.793 | 0.953 | 0.905 - 1.001 | 0.841 | 0.968 | 0.928 - 1.008 |
| | Healthy | 0.150 | 1 | | 0.301 | 1 | |
| | Overweight | 1 | **0.901** | **0.860 - 0.943** | 0.763 | 0.974 | 0.935 - 1.012 |
| | Obese | 0.914 | **0.941** | **0.896 - 0.987** | 0.246 | 1.003 | 0.959 - 1.047 |
| Smoke | Daily | 0.294 | 1 | | 0.855 | 1 | |
| | Never | 0.503 | 0.986 | 0.937 - 1.034 | 0.486 | 1.022 | 0.978 - 1.066 |
| | Sometimes | 0.723 | 0.971 | 0.922 - 1.020 | 0.262 | 1.036 | 0.991 - 1.081 |
| Snuff | Daily | 0.406 | 1 | | 0.596 | 1 | |
| | Never | 0.414 | 0.999 | 0.949 - 1.049 | 0.310 | 1.017 | 0.973 - 1.060 |
| | Sometimes | 0.900 | 0.962 | 0.915 - 1.010 | 0.831 | 0.986 | 0.947 - 1.025 |
| Alcohol | ≥ 2 per month | **0.035** | 1 | | 0.434 | 1 | |
| | ≤1 month | 0.806 | **0.933** | **0.885 - 0.980** | 0.602 | 0.991 | 0.948 - 1.034 |
| | Never | 0.739 | **0.938** | **0.890 - 0.986** | 0.323 | 1.006 | 0.964 - 1.049 |
| Physical activity[b] | Light | 0.301 | 1 | | 0.089 | 1 | |
| | None | 0.994 | **0.930** | **0.886 - 0.975** | 0.803 | **0.952** | **0.914 - 0.990** |
| | Medium | **0.008** | 1.053 | 0.999 - 1.107 | 0.267 | 0.982 | 0.941 - 1.023 |
| | Hard | 0.883 | **0.959** | **0.913 - 1.005** | 0.817 | **0.951** | **0.913 - 0.989** |
| Hormonal contra-ceptives[c] | Non-user | 0.444 | 1 | | 0.494 | 1 | |
| | Progestin | 0.369 | 1.126 | -0.230 - 2.482 | 0.392 | 1.239 | -0.576 - 3.054 |
| | Low Estrogen | 0.430 | 1.024 | -0.414 - 2.264 | 0.475 | 1.046 | -0.840 - 2.932 |
| | High Estrogen | 0.476 | 0.940 | -0.546 - 2.425 | 0.483 | 1.027 | -0.862 - 2.916 |
| P-values from comparison between random network against a random network with only that particular | | | | | | | |

category. Participants with missing values are excluded from the analysis. Statistically significant values highlighted in bold.
Relative risk and 95% confidence interval (95% CI) from univariable logistic regression analysis.

[a] BMI = body mass index.by kg/m$^2$. Underweight = <18.5; Healthy = 18.5-24.9; Overweight = 25.0-29.9; $\geq$ 30.0
[b] Physical activity: None = reading, watching TV, or other sedentary activity; Low level = walking, cycling, or other forms of exercise at least 4 hours a week; Medium level = participation in recreational sports, heavy outdoor activities with minimum duration of 4 hours a week; High level = Participation in heavy training or sports competitions regularly several times a week.
[c] Hormonal contraceptives: Non-user = No current use of hormonal contraceptives (women only); Progestin-only = Use of hormonal contraceptives with progestin (Cerazette, Nexplanon, Depo-provera, Implanon); Combination contraceptives low estradiol = Use of hormonal contraceptives with progestin and ethinyl estradiol less than or equal to 20µg (Mercilon, Yasminelle, Loette 28, Nuvaring). Combination contraceptives high estradiol = Use of hormonal contraceptives with progestin and ethinyl estradiol greater than or equal to 30µg (Marvelon, Yasmin, Microgynon, Oralcon, Diane, Synfase, Evra, Zyrona). Women taking contraceptives, but who were unable to recognize the brand were removed from the analysis

**Table 2 Correlation between host risk factors and *Staphylococcus aureus* carrier status.** Fit Futures 1 (N = 1038). Adapted multivariable linear autocorrelation model.

|  | Estimate[a] | Std Error | P-value |
| --- | --- | --- | --- |
| **Direct culture** | | | |
| ρ | 0.048 | 0.011 | **<0.001** |
| Sex | -0.048 | 0.028 | **0.0016** |
| Study program | 0.043 | 0.022 | 0.0542 |
| BMI[b] | 0.107 | 0.018 | **<0.001** |
| Smoke | -0.012 | 0.027 | 0.650 |
| Snuff | -0.001 | 0.020 | 0.968 |
| Alcohol | 0.038 | 0.021 | 0.066 |
| Physical activity | 0.033 | 0.012 | **0.008** |
| **Enrichment culture** | | | |
| ρ | 0.060 | 0.010 | **<0.001** |
| Sex | -0.017 | 0.030 | 0.578 |
| Study program | 0.087 | 0.024 | **<0.001** |
| BMI[b] | 0.085 | 0.019 | **<0.001** |
| Smoke | -0.019 | 0.029 | 0.517 |
| Snuff | 0.001 | 0.022 | 0.950 |
| Alcohol | 0.070 | 0.022 | **0.002** |
| Physical activity | 0.046 | 0.014 | **<0.001** |

Significant values highlighted in bold.

[a] Only estimates for the total model are valid. Beta estimates for individual host factors cannot be interpreted.

[b] BMI = body mass index

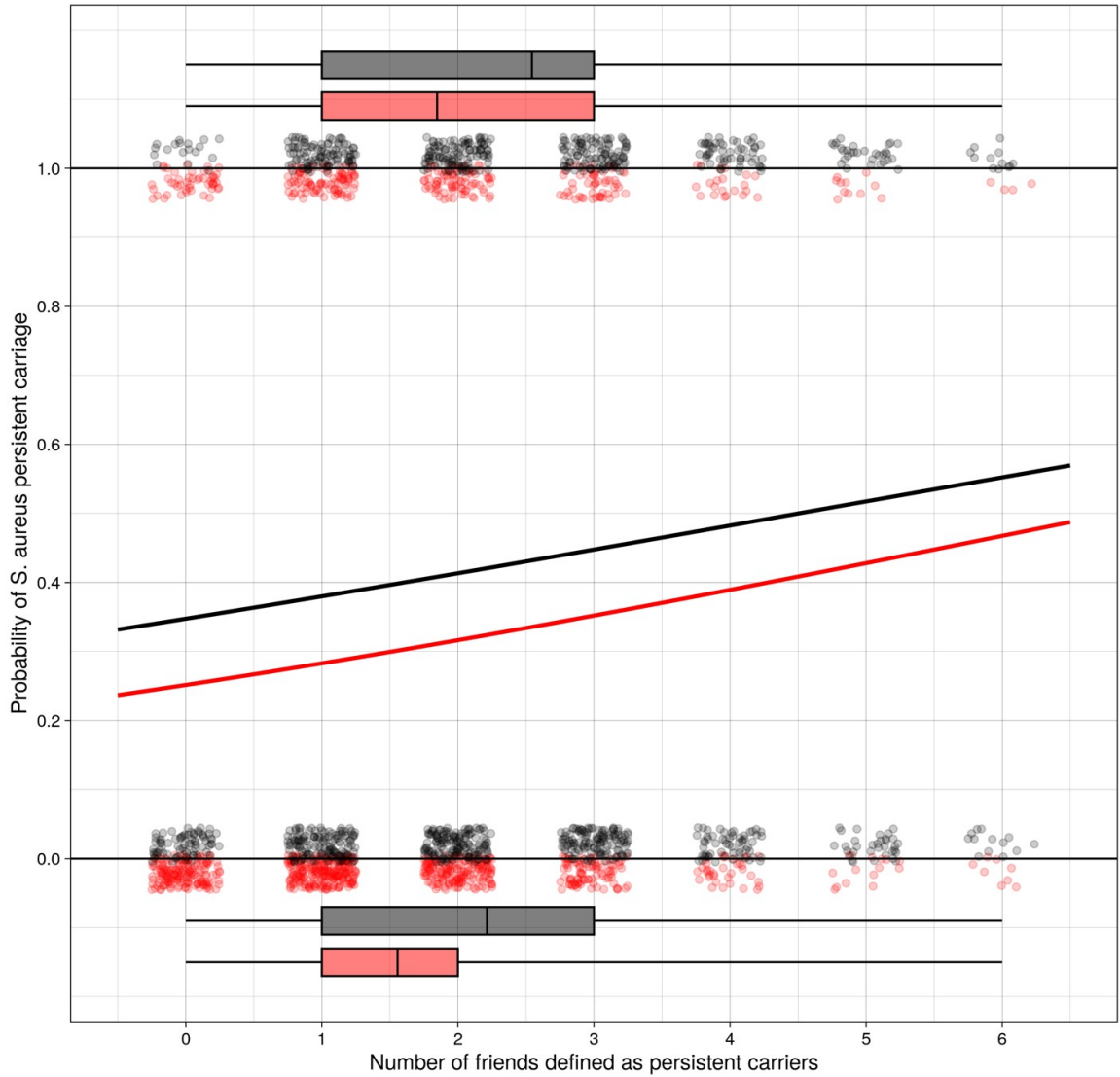

**Figure 3 Probability of Staphylococcus aureus persistent nasal carriage with respect to friends' carrier status.** Univariable logistic regression analysis. Fit Futures 1 (n=1038). Red color represents persistent nasal carriage defined by direct culture. Black color represents persistent nasal carriage defined by enrichment culture

The scatterplots show the distribution of persistent nasal carriers (along Y = 1) and non-carriers (along Y = 0) for sub-populations of students having from 0 to 6 *S. aureus* positive friends.

**Boxplots show the mean (middle line) and interquartile range (box limits) of S. aureus positive friends for persistent nasal carriers (at the top of the diagram) and non-carriers (at the bottom of the diagram). Outliers with more than 6 S. aureus positive friends are excluded from the figure, but did not affect the result (N = 3). For more information see Supplementary Table 6 and 7.**

## Discussion

In the present study, we demonstrate that social network is associated with *S. aureus* persistent carrier status and *spa*-type in a youth population. This is, to our knowledge, the first study to analyze transmission of *S. aureus* in a general population. We show that your own probability of being a persistent carrier correlates with the number of close friends that are colonized with *S. aureus*. The autocorrelation analysis shows a 5-6% increased probability of *S. aureus* carriage with each additional *S. aureus* carrier friend, with comparable results of 3·4-3·7% from the logistic regression analysis. We also show that friends tend to have the same *spa*-type, supporting that the social network effect is partly driven by direct transmission of *S. aureus*. Our results coincide with former research that has demonstrated comparable results in different cohorts (13, 14).

We analyzed different types of networks and found an association between transmission of *S. aureus* in the social network where participants confirmed they had physical contact (direct culture: p=0·07; enrichment culture: p=0·04). This may indicate that social contact is a key pathway for the spread of *S. aureus* in the community (19), which is in line with former studies of transmission of *S. aureus* through a social network (14). Being together at school was also significantly associated with *S. aureus* transmission. We did not find any significant spread of *S. aureus* among students being together at home and in sports.

There is also a substantial social dimension for several of the known host risk factors for *S. aureus* carriage, which suggests that social network effects may have contributed to associations observed in former studies. In a univariable logistic regression model, there was an increased risk of 3·4-3·7% of being a persistent carrier when increasing your friend circle by one *S. aureus* positive friend. A similar analysis (autocorrelation model) adjusted for host risk factors, showed an increase of 4·8-6·0%. The difference between the logistic regression model and the autocorrelation model may partly represent the effect of the individual risk factors.

In our population, male students had a higher prevalence of *S. aureus* persistent carriage compared to females which corresponds with previous studies (9,20). The social network analysis demonstrates that female sex is the predominant social risk factor for carriage due to more relationships among females. This may substantiate the hypothesis of sex as a true biological risk factor for *S. aureus* carriage, as the male population have a higher prevalence of carriage while the relative risk of transmission is lower compared to the female population.

We also demonstrate increased transmission of *S. aureus* among students engaged in medium level physical activity in leisure time compared to among those with sedentary leisure time. A former study showed an increased risk of *S. aureus* carriage in athletes doing contact sport (20). Many of the physical activities in youths are contact sports or training in close counters. In our population there was a higher percentage of females doing medium physical activity compared to males (women = 27% and men = 23%). The increased risk of spread related to medium physical activity could therefore be partly attributed to the observed sex differences in risk of transmission.

The use of alcohol more than twice a month was a social factor associated with carriage of *S. aureus*. This may reflect increased social contact with multiple friends at parties and social gatherings. Participants consuming alcohol more than two times a month had a higher number of friends than participants consuming less or no alcohol. We do not have information about the amount of alcohol consumed, and the alcohol variable is therefore lacking some precision. We also have some outliers that may have affected the results.

The autoregression model suggests an association between BMI and transmission of *S. aureus* in addition to sex, alcohol use, physical activity, and study program. The effect of friendship density may partly underlie this association, as students with normal BMI had more friends.

Excluding older outliers above 20 years (n=36) from the network analysis did not affect the results, and all participants were therefore kept in the analysis. Another bias in this study is the precision of the interview which was the basis for the social networks. None of the questions provide information about the type or amount of physical contact, which will give a bias of unknown size and direction. The social networks were constructed by self-reported information on social contacts the last week prior to the study, and this could be misrepresenting of the participants' social contact over longer periods of time. We therefore asked all participants to score the representativeness of the nominated friends, and 76 % of the participants claimed a score of five or above (on a scale from one to ten). We therefore believe the representativeness of the nominations to be high (Supplementary Figure 5).

We had complete *spa*-type data only from throat isolates, while nasal carriage is generally considered as the most clinically relevant phenotype. In a validation study among 100 participants with *S. aureus* isolated from both the two nasal and the two throat swab cultures in FF1, 82 participants had the same *spa*-type in both sites (data not shown). We therefore believe that our findings from social network analysis based on *spa*-type of throat isolates are representative also for transmission of nasal *S. aureus*. Another limitation is that we had 10 invalid nasal samples for the first swab, and 51 invalid samples for the second swab. These were reclassified as negative for *S. aureus*. Because of the analysis of social network, we believe that it would have introduced a larger bias excluding parts of the social network compared to the bias of including potentially misclassified samples.

Our analysis was modeled by using one time point, while interviews with the different participants were conducted at multiple timepoints. Most participants nominated friends who had the same attendance date as themselves, e.g., from their own school class (Supplementary Table 8). Furthermore, persistent nasal carriage is a relatively stable phenotype (17), and we therefore assume that time will not affect the present analysis.

In summary, our data from a general youth population supports social effects on *S. aureus* carriage and that these result from both direct social transmission and shared lifestyle risk factors for carriage among friends. We demonstrate relationships between different social networks (i.e., overall, physical contact, school) and *S. aureus* persistent carriage and specific *spa*-types. We can also show that risk of transmission differs by host lifestyle factors. The male predominance in carriage is determined by sex-specific predisposing host characteristics, as social interactions among men are weak drivers of transmission compared to women. More studies are needed to further evaluate the interplay between the social environment and host risk factors in *S. aureus* carriage.

## Declaration of interest

All authors state no conflict of interest.

## Acknowledgements

We are grateful for the contribution by the participants of the Fit Futures 1 study. We also thank the Clinical Research Department at the University Hospital of North Norway, Tromsø, for data collection. We also want to thank the Department of Medical Microbiology and the Department of Community Medicine at the Faculty of Health Sciences.

## Funding

The Northern Norway regional Health Authorities (grant number HNF1457-19) funded this study.

## Data sharing

The data that support the findings of this study are available from The Fit Futures study but restrictions apply to the availability of these data, which were used under license for the current study, and so are not publicly available. Data are however available from the authors upon request and with permission of The Fit Futures study. Proposals for data should be directed to fitfutures@uit.no. Statistical analysis and consent form will be available on request. Proposals should be directed to dina.b.stensen@uit.no.

## Authors' contributions

Anne-Sofie Furberg, Christopher Sievert Nielsen, Gunnar Skov Simonsen and Lars Ailo Bongo contributed with the conceptualization and design of the work. Anne-Sofie Furberg and Lars Ailo Bongo supervised the work. Johanna UE Sollid performed microbiological analysis of nasal and throat samples. Rafael A. Nozal Canadas contributed with statistical analysis and statistical methods. Karina Olsen, Lars Småbrekke, Kristian Svendsen, Dina Stensen and Anne Merethe Hanssen contributed to interpretation of data. Dina B. Stensen and Rafael A. Nozal Canadas wrote the original draft. All authors reviewed and approved the final manuscript. Rafael A. Nozal Canadas and Lars Ailo Bongo verify the underlying data.

# Supplementary Material

## Table of contents

# Statistical background

The use of Network analysis has increased exponentially over the last few years. In this brief introduction we include the statistical background on random graph analysis and autocorrelation networks models, and some included references to provide a deeper understanding of the topic.

In statistics, a general rule to solve problems is to find all possibilities and compare all the scenarios in which something happens against all scenarios in which something does not happen. Such a ratio will give you the probability of something occurring. In our case, it is impossible to compare random graphs with all possible random graphs to get the real probability. Such calculations are unattainable, and it is necessary to constrain the amount of possible random graphs based on some assumptions (1). The constraints added to the random graph will give a model which is similar enough to reality. In our case, we use the same frequency tables with a network with the same topology as constriction. We also assume that high contagiousness would cluster positives together with positives, and negatives together with negatives.

In this context, identical topology means having the same nodes (participants) and same edges (relationships) as the original network; but each node has randomly assigned attributes based on the probability distribution of each category (i.e., *S. aureus* persistent carrier status is assigned randomly to each node, following an arbitrary 30% prevalence probability, instead of using the original value).

Our bootstrapping (2) consists of counting how many relationships connect two nodes with the same attributes in our network (i.e., persistent carrier with persistent carrier or same *spa*-types) in 1000 simulations. This gives us a distribution of 1000 values (with a mean and standard deviation) which we can compare to the real number of homophilic relationships in our network. We then perform simple hypothesis testing like t-test, where we consider a p-value of 0.05 or less to be statistically significant. As the numbers of these tests are low, there is no need for p-value correction for false positives.

Further, we can use the random network average number of relationships that we just created to compare with our network. We can create, again, a random network with the same topology, but using the conditional probability for each host factor independently. In this way we can check how much each of the categories deviates with respect to each other, and we can identify which category has higher or lower risk for the outcome variable.

Network autocorrelation models (3) are a special case of autoregression analysis in time series (4) where we want to find how much influence your neighbors (typically your social network of friends) have over you. We aim to find the ρ coefficient in the formula:

$$Y^{(t+1)} = \rho W_n Y^{(t)} + X\beta + \varepsilon$$

W is a weighted matrix (typically normalized to 1) indicating which neighbors have influence over you, X is your explanatory variable (in our case, sex, BMI, smoke, and so on), β is a vector of coefficients (similar to linear regression) and ε is a random noise vector. Y is your dependent variable vector (in our case persistent carrier status), which over time (t), will converge to a common value. The ρ coefficient represents how much you are following the pressure of your neighbor influence and ranges typically from 0 to infinity, although negative values are also valid depending on your context. A value close to 0 would mean that you are completely ignoring your neighbors and the explanatory variables really do not have any influence on you. Positive values indicate that people can exert influence over you, and in our case, your friends will increase your risk of being a carrier.

Significant negative values would indicate that you dislike your neighbors so much, that you will do the complete opposite of what he tells you to do. In our case, this is not a valid case for the ρ coefficient as you cannot protect someone from carriage, you simply will not transmit the bacteria. However, for the explanatory variables, negative values of ρ coefficient are valid in our case because we encode categorical variables with dummy variables.

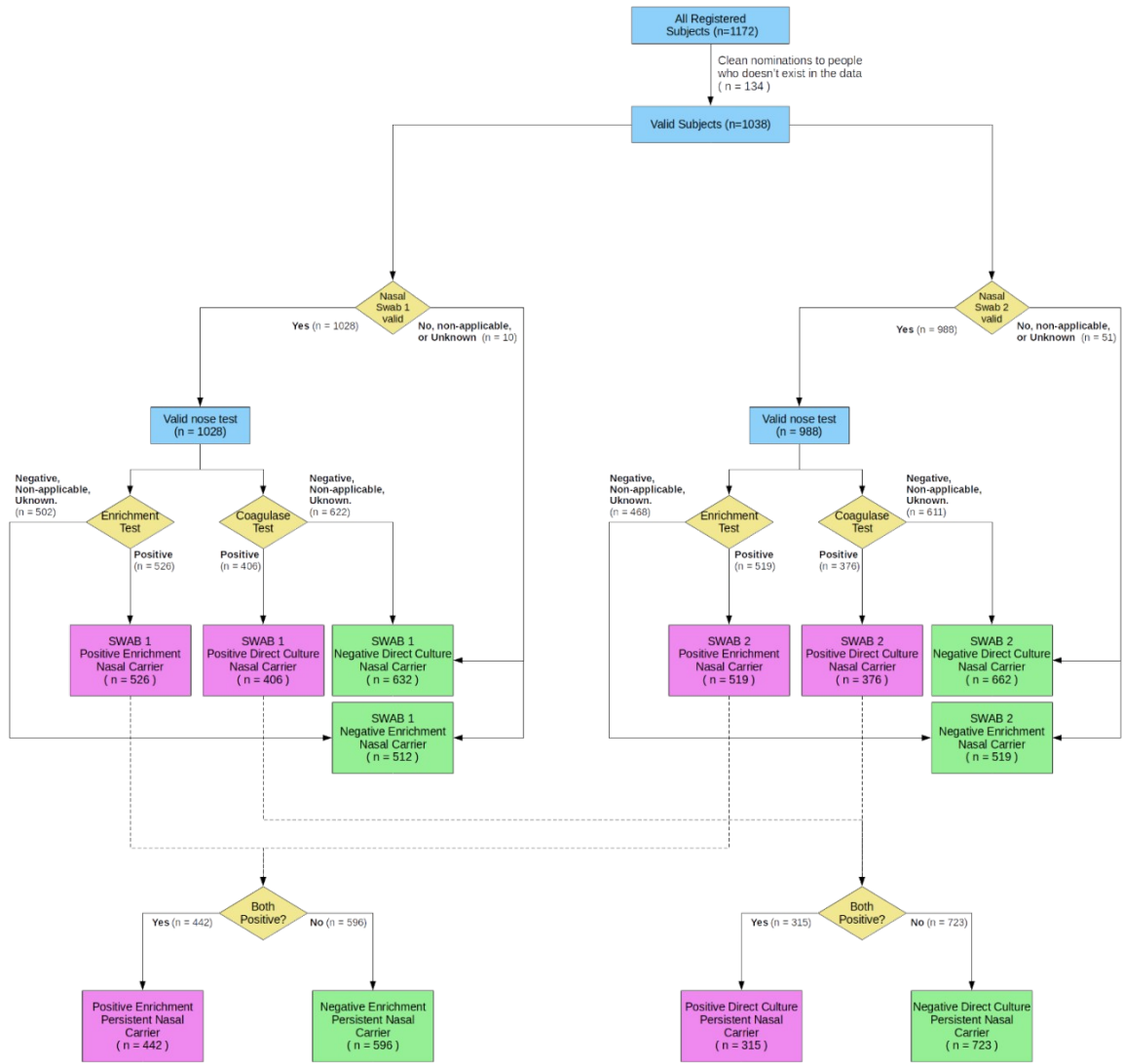

**Supplementary Figure 1 Flowchart with inclusion criteria for definitions of *Staphylococcus aureus* persistent nasal carriage.** The Fit Futures 1 study (N = 1038). Direct culture (bottom right) and enrichment culture (bottom left).

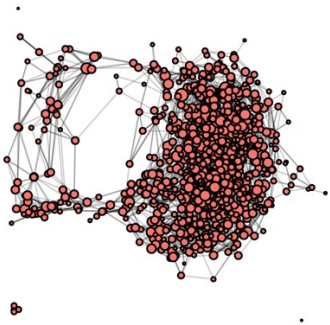
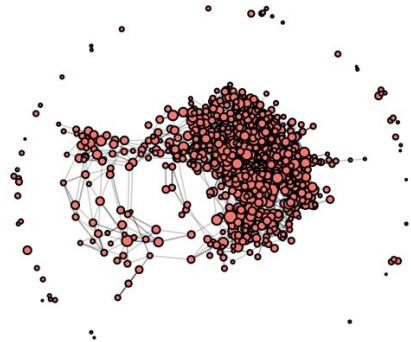
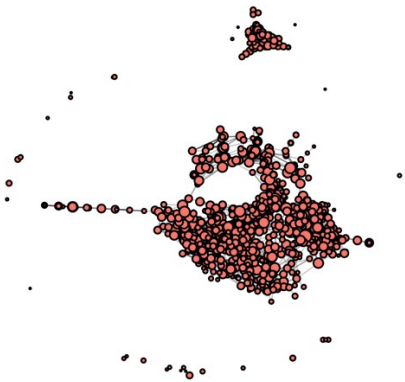
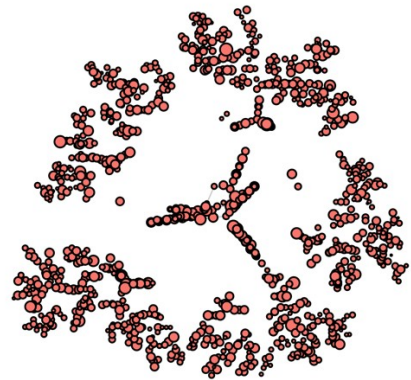
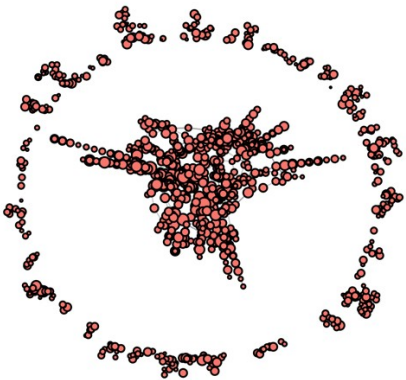
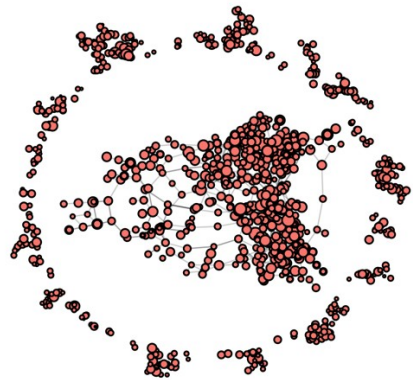

**Supplementary Figure 2 Overview of the different social networks.** The Fit Futures 1 study (N = 1038). From top to bottom and left to right: overall social network, physical contact, together at school, together in sports, together at home, and together in other settings. Each node represents a student. Each edge represents an undirected nomination (connection). The size of the node is proportional to the number of connections.

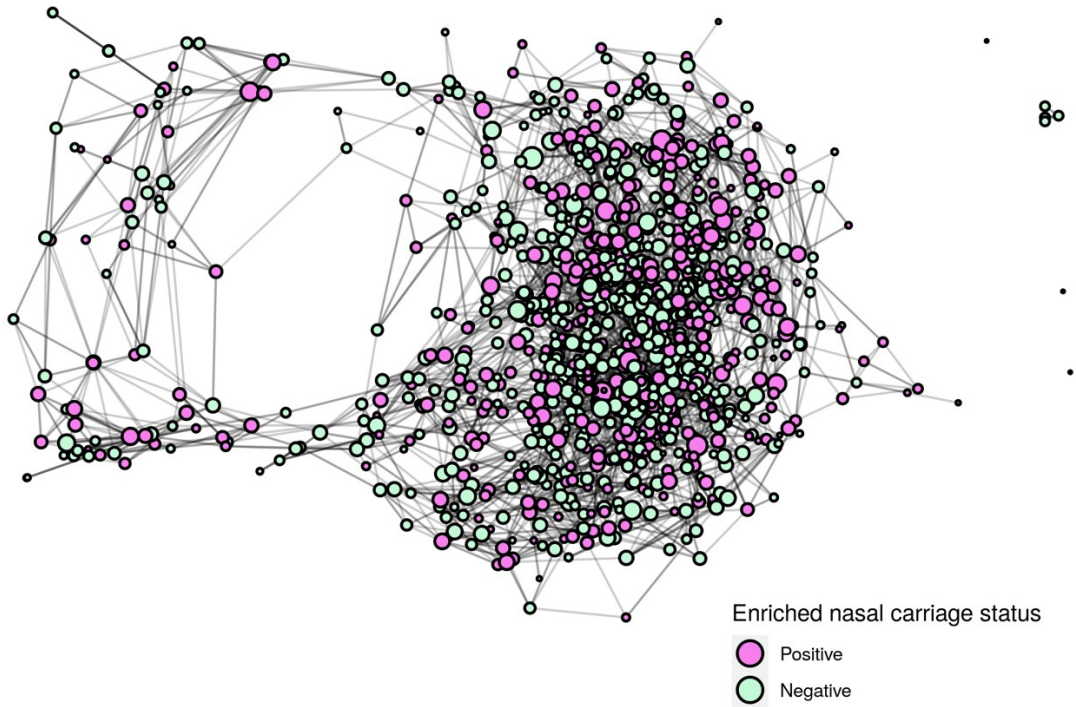

**Supplementary Figure 3 Overall network.** The Fit Futures 1 study (N = 1038). Staphylococcu aureus persistent nasal carriage status determined by enrichment culture is highlighted for each student (Positive = S. aureus detected in two nasal swab samples; Negative = S. aureus detected in one or none of two nasal swab samples). Node size is proportional to the number of connections (undirected friendship).

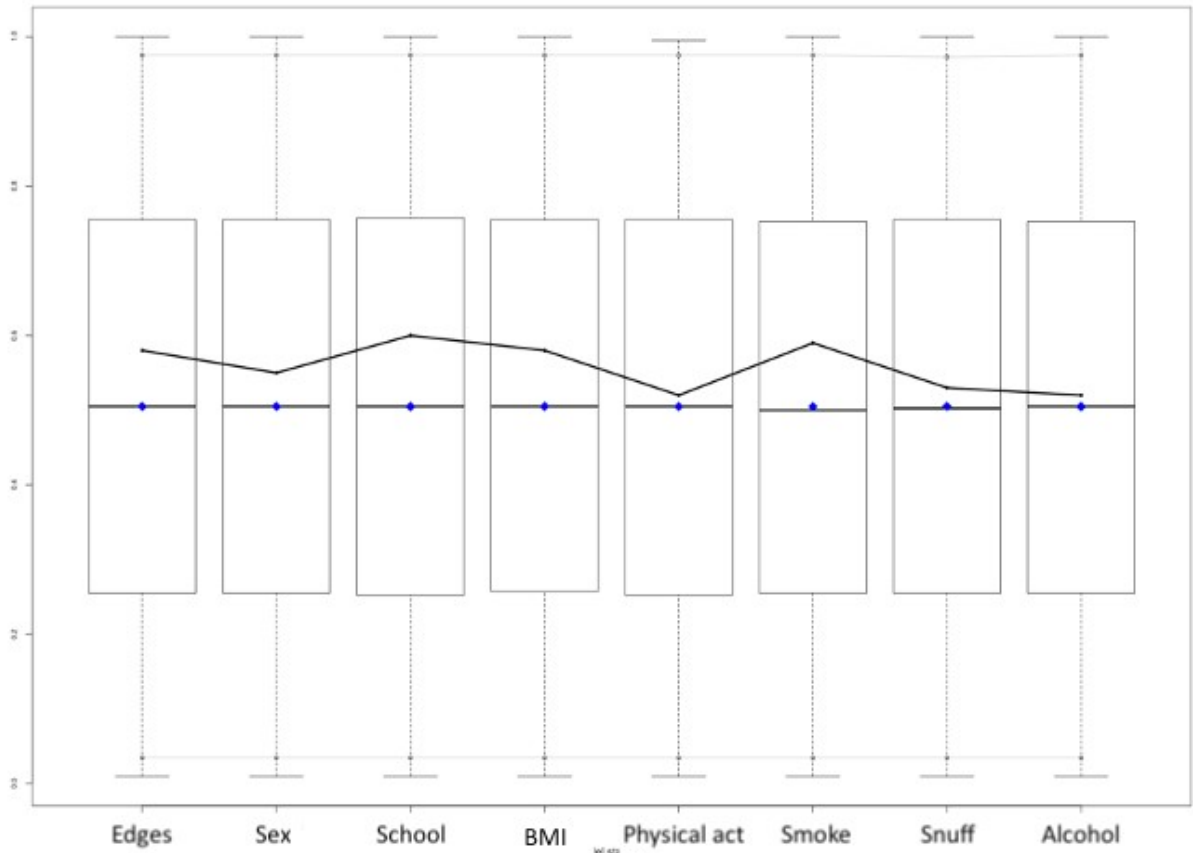

**Supplementary Figure 4 Goodness of fit for the ERGM (Exponential Random Graph Model) analysis.** The Fit Futures 1 study (N = 1038). Y-axis = proportion of statistics, X-axis = model statistics.

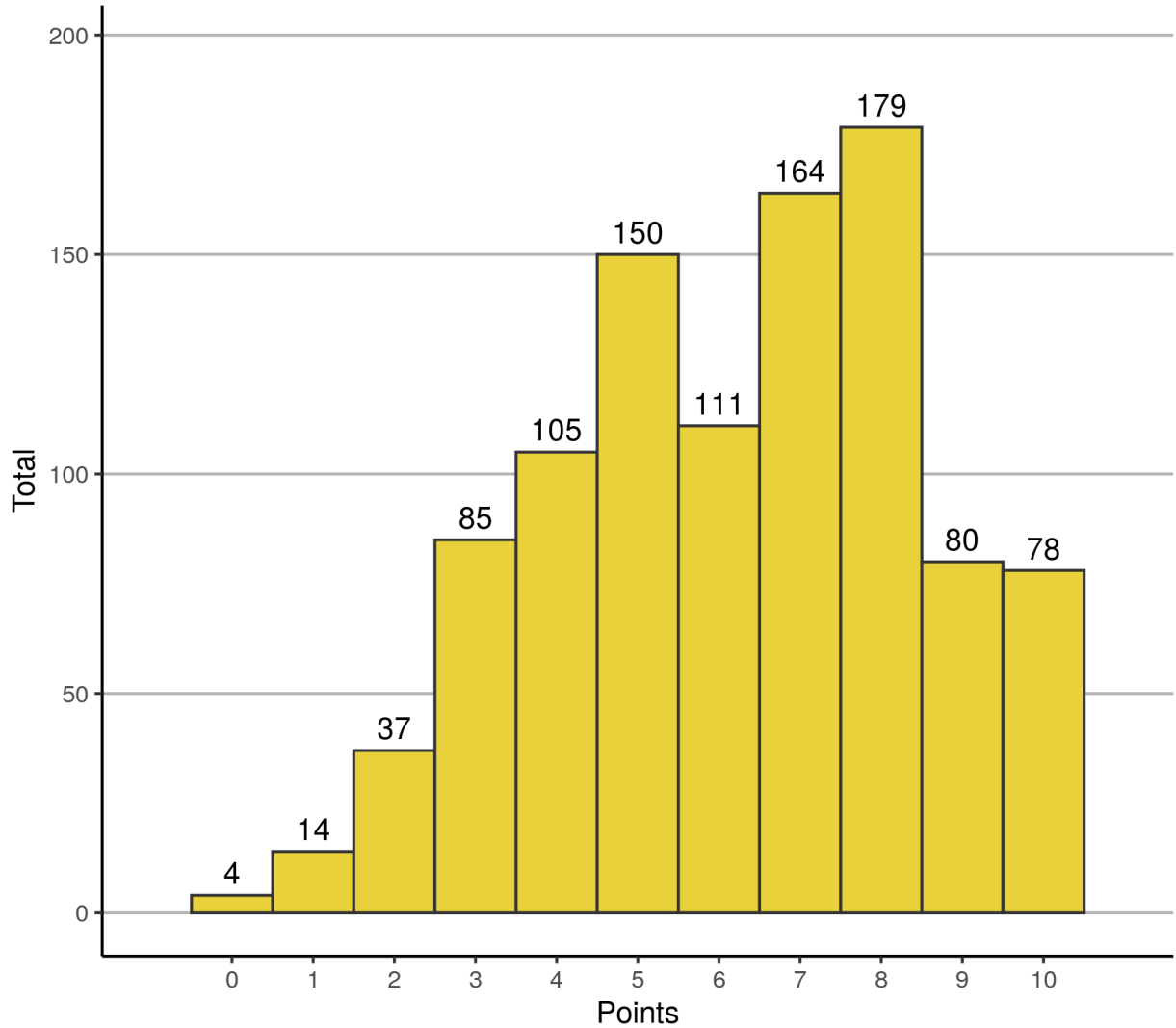

**Supplementary Figure 5 Histogram of representativeness of the social network.** The Fit Futures 1 study (N = 1038). 0 being not representative at all, and 10 being a perfect overview. For specific networks, having a mean score from top to bottom of school (6.61), other (6.52), physical (6.42) overall (6.29) sports (6.24) and home (6.13) (information not included in figure).

**Supplementary Table 1 Characteristics of the study population by *Staphylococcus aureus* persistent nasal carriage determined by direct and enrichment culture.** The Fit Futures 1 study (N = 1038).

|  | Direct culture | | | Enrichment culture | | |
|---|---|---|---|---|---|---|
|  | Positive[d] | Negative[d] | Prevalence | Positive[d] | Negative[d] | Prevalence |
| Sex |  |  | < 0.001 |  |  | < 0.001 |
| Male | 193 | 337 | 36.4 % | 255 | 275 | 48.1 % |
| Female | 122 | 386 | 24.0 % | 187 | 321 | 36.8 % |
| Study program |  |  | 0.99 |  |  | 0.08 |
| General | 118 | 272 | 30.3 % | 163 | 227 | 41.8 % |
| Sports | 31 | 73 | 29.8 % | 55 | 49 | 52.9 % |
| Vocational | 166 | 378 | 30.5 % | 224 | 320 | 41.2 % |
| Smoking |  |  | 0.93 |  |  | 0.48 |
| Daily | 14 | 34 | 29.2 % | 24 | 24 | 50.0 % |
| Sometimes | 59 | 129 | 31.4 % | 76 | 112 | 40.4 % |
| Never | 236 | 546 | 30.2 % | 333 | 449 | 42.6 % |
| Snuff use |  |  | 0.79 |  |  | 0.30 |
| Daily | 73 | 172 | 29.8 % | 107 | 138 | 43.7 % |
| Sometimes | 43 | 88 | 32.8 % | 63 | 68 | 48.1 % |
| Never | 192 | 450 | 29.9 % | 263 | 379 | 41.0 % |
| BMI category[a] |  |  | 0.21 |  |  | 0.22 |
| < 18.5 kg/m2 | 35 | 75 | 31.8 % | 55 | 55 | 50.0 % |
| 18.5-<25 kg/m2 | 201 | 509 | 28.3 % | 289 | 421 | 40.7 % |
| 25-<30 kg/m2 | 54 | 93 | 36.7 % | 68 | 79 | 46.3 % |
| ≥30 kg/m2 | 22 | 45 | 32.8 % | 27 | 40 | 40.3 % |
| Physical activity[b] |  |  | 0.15 |  |  | 0.07 |
| None | 80 | 149 | 34.9 % | 107 | 122 | 46.7 % |
| Light | 99 | 239 | 29.3 % | 129 | 209 | 38.2 % |
| Medium | 67 | 192 | 25.9 % | 105 | 154 | 40.5 % |
| Hard | 63 | 131 | 32.5 % | 93 | 101 | 47.9 % |
| Alcohol intake | | | 0.32 | | | 0.780 |
| Never | 88 | 192 | 31.4 % | 115 | 165 | 41.1 % |
| <= 1 Month | 134 | 286 | 31.9 % | 183 | 237 | 43.6 % |
| ≥2 Month | 86 | 232 | 27.0 % | 134 | 184 | 42.1 % |
| Hormonal contraceptives[c] | | | 0.76 | | | 0.68 |
| Non-user | 78 | 249 | 23.9 % | 121 | 206 | 37.2 % |
| Progestin only | 3 | 17 | 15.0 % | 5 | 15 | 25.0 % |
| Combination contraceptives, low estradiol | 12 | 38 | 24.0% | 19 | 31 | 38.0 % |
| Combination contraceptives, high estradiol | 26 | 73 | 27.1 % | 39 | 60 | 39.4 % |

[a] BMI = body mass index

[b] Physical activity in leisure time: None = reading, watching TV, or other sedentary activity; Low level = walking, cycling, or other forms of exercise at least 4 hours a week; Medium level = participation in recreational sports, heavy outdoor activities with minimum duration of 4 hours a week; High level = Participation in heavy training or sports competitions regularly several times a week.

[c] Hormonal contraceptives: Non-user = No current use of hormonal contraceptives (women only); Progestinonly = Use of hormonal contraceptives with progestin (Cerazette, Nexplanon, Depo-provera, Implanon); Combination contraceptives, low estradiol = Use of hormonal contraceptives with progestin and ethinyl estradiol less than or equal to 20μg (Mercilon, Yasminelle, Loette 28, Nuvaring). Combination contraceptives, high estradiol = Use of hormonal contraceptives with progestin and ethinyl estradiol greater than or equal to 30μg (Marvelon, Yasmin, Microgynon, Oralcon, Diane, Synfase, Evra, Zyrona). Women taking contraceptives, but who were unable to recognize the brand were removed from the analysis.

[d] Positive = two consecutive nasal swab cultures positive for *Staphylococcus aureus* Negative = one or none of two consecutive nasal swab cultures positive for *Staphylococcus aureus*

**Supplementary Table 2 ERGM (Exponential Random Graph Model) analysis of relationships within groups of participants with the same characteristics.** The Fit Futures 1 study (N = 1038).

|  | Homophily (%) | Estimate (logit) | Std Error | P-value |
|---|---|---|---|---|
| Edges | -- | -8.41 | 0.08 | **< 0.001** |
| Sex | 84.05 | 1.47 | 0.05 | **< 0.001** |
| School | 87.85 | 2.16 | 0.06 | **< 0.001** |
| BMI[a] | 54.23 | 0.18 | 0.04 | **< 0.001** |
| Smoke | 68.06 | 0.22 | 0.04 | **< 0.001** |
| Snuff | 57.71 | 0.31 | 0.04 | **< 0.001** |
| Alcohol | 45.26 | 0.42 | 0.04 | **< 0.001** |
| Physical activity | 40.22 | 0.43 | 0.04 | **< 0.001** |

[a] BMI = body mass index

**Supplementary Table 3 The most prevalent *spa*-types for *Staphylococcus aureus* throat carriage.** The Fit Futures 1 study (N = 746). Only persistent carriers are shown. The plots are the results for the direct culture (above) and for enrichment culture (below).

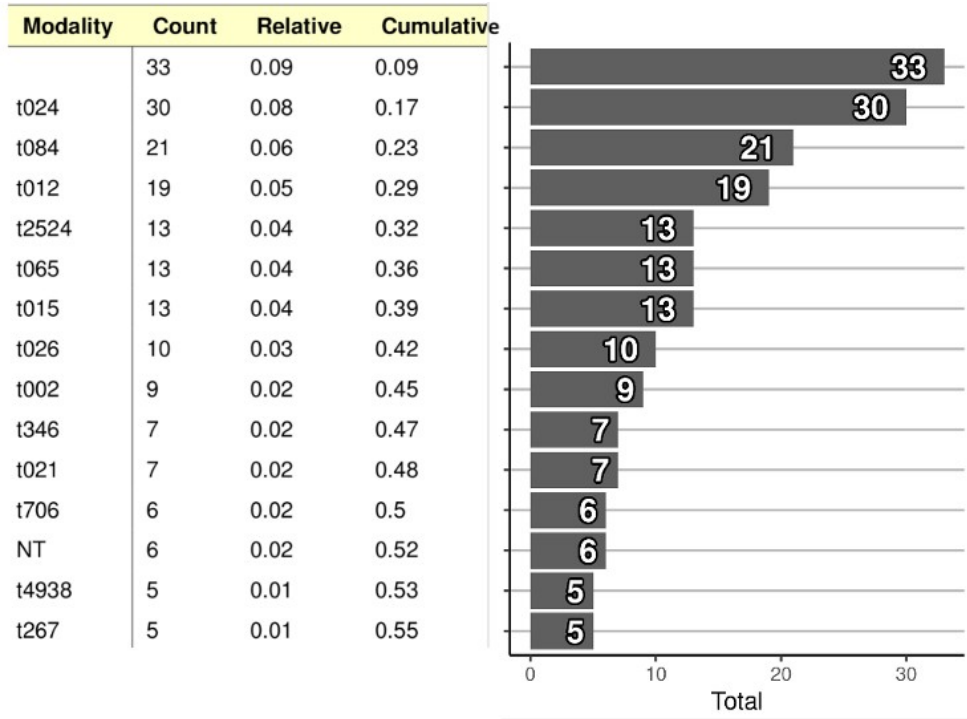

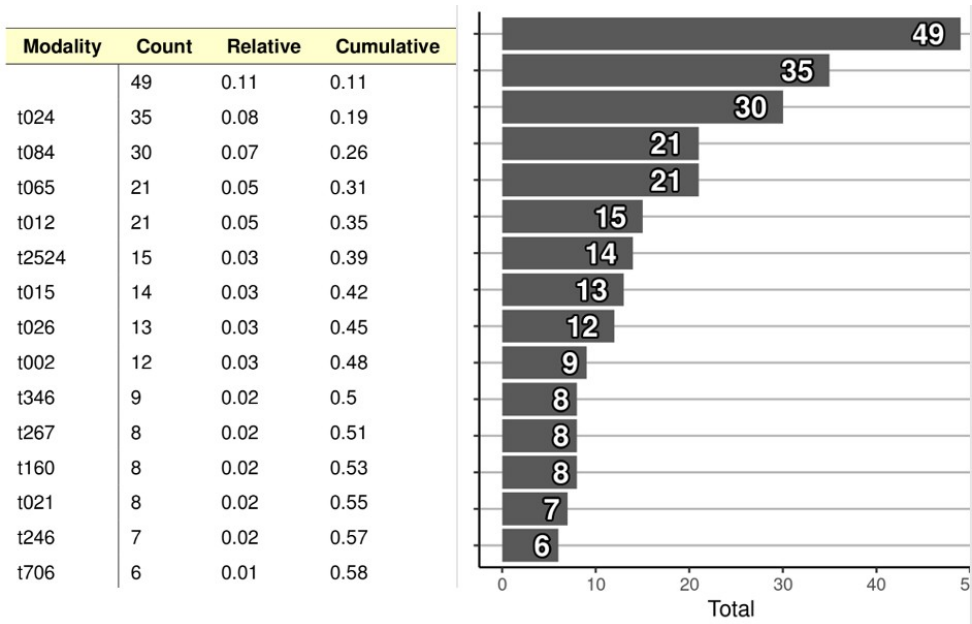

**Supplementary Table 4 Detailed summary of 1000 simulations for each social network.** The Fit Futures 1 study (N = 1038).

| Network | Total Relationships | Equal relationships | MIN | Q1 | Median | Q3 | MAX | SD | Direct culture P-value |
|---|---|---|---|---|---|---|---|---|---|
| Overall | 3767 | 2260 | 2012 | 2136 | 2177 | 2214 | 2353 | 57 | 0.07 |
| Physical | 2823 | 1698 | 1492 | 1596 | 1628 | 1658 | 1756 | 45 | 0.06 |
| School | 2979 | 1814 | 1559 | 1687 | 1718 | 1747 | 1866 | 46 | **0.02** |
| Sports | 598 | 365 | 285 | 333 | 345 | 357 | 404 | 17 | 0.12 |
| Home | 1247 | 731 | 644 | 703 | 720 | 737 | 812 | 25 | 0.34 |
| Others | 1095 | 663 | 567 | 616 | 632 | 648 | 705 | 22 | 0.08 |
| Network | Total relationships | Equal relationships | MIN | Q1 | Median | Q3 | MAX | SD | Enrichment culture P-value |
| Overall | 3767 | 2013 | 1784 | 1899 | 1926 | 1953 | 2040 | 40 | **0.02** |
| Physical | 2823 | 1502 | 1339 | 1418 | 1442 | 1465 | 1576 | 34 | **0.04** |
| School | 2979 | 1610 | 1401 | 1499 | 1524 | 1548 | 1647 | 36 | **0.01** |
| Sports | 598 | 314 | 257 | 296 | 306 | 315 | 367 | 15 | 0.29 |
| Home | 1247 | 644 | 570 | 623 | 638 | 652 | 717 | 22 | 0.39 |
| Others | 1095 | 588 | 507 | 545 | 558 | 571 | 625 | 19 | 0.06 |
| Network | Total relationships | Equal relationships | MIN | Q1 | Median | Q3 | MAX | SD | *Spa*-type P-value |
| Overall | 1948 | 136 | 20 | 45 | 51 | 58 | 90 | 9.6 | **< 0.001** |
| Physical | 1459 | 111 | 16 | 33 | 38 | 44 | 84 | 8.0 | **< 0.001** |
| School | 1539 | 100 | 15 | 35 | 41 | 46 | 76 | 8.2 | **< 0.001** |
| Sports | 335 | 21 | 0 | 7 | 9 | 12 | 22 | 3.7 | **< 0.001** |
| Home | 664 | 63 | 4 | 14 | 17 | 21 | 38 | 5.1 | **< 0.001** |
| Others | 563 | 45 | 4 | 12 | 15 | 18 | 30 | 4.5 | **< 0.001** |

Columns 4-9[4] contain the simulation summary statistics of the 1000 simulation result, in order, the minimum value of same-to-same relationships, first quartile, median rounded to the nearest integer, third quartile, maximum value, and standard deviation rounded to the nearest integer. The last column is the result of applying a t-test with the equal relationship against a distribution formed with the average of the 1000 simulations, and the standard deviation of the 1000 simulations. Significant p-values are highlighted in bold.

**Supplementary Table 5  Average popularity in the overall network for each host risk factor.**
The Fit Futures 1 study (N = 1038).

|  | Average Popularity [a] (3.62) | Relative physical isolation [b] (%) | Relative frequency all (%) |
|---|---|---|---|
| **Sex** | **0.008** | | |
| Male | 3.46 | 70 | 51.1 |
| Female | 3.81 | 30 | 48.9 |
| **BMI-category [c]** | **0.001** | | |
| < 18.5 kg/m2 | 3.61 | 9.23 | 10.60 |
| 18.5-<25 kg/m2 | 3.72 | 62.31 | 68.40 |
| 25-<30 kg/m2 | 3.63 | 14.62 | 14.16 |
| > 30 kg/m2 | 2.64 | 13.08 | 6.45 |
| **Smoking** | **0.003** | | |
| Daily | 2.75 | 10 | 4.62 |
| Sometimes | 3.90 | 13.85 | 18.11 |
| Never | 3.60 | 72.31 | 75.34 |
| **Snuff use** | **0.003** | | |
| Daily | 3.82 | 19.23 | 23.60 |
| Sometimes | 4.05 | 7.69 | 12.62 |
| Never | 3.45 | 69.23 | 61.85 |
| **Study program** | **0.004** | | |
| General | 3.83 | 33.08 | 37.57 |
| Sports | 3.95 | 5.38 | 10.02 |
| Vocational | 3.43 | 62.31 | 52.41 |
| **Physical activity [d]** | 0.254 | | |
| None | 3.45 | 29.23 | 22.06 |
| Light | 3.56 | 30.77 | 32.56 |
| Medium | 3.73 | 21.54 | 24.95 |
| Hard | 3.80 | 14.62 | 18.69 |
| Alcohol intake | **< 0.001** | | |
| Never | 3.05 | 41.54 | 26.97 |
| <= 1 Month | 3.76 | 30.77 | 40.46 |
| > 2 Month | 3.93 | 23.85 | 30.64 |
| Direct culture persistent carriage | 0.347 | | |
| Positive | 3.72 | 30 | 30.35 |
| Negative | 3.59 | 70 | 69.65 |
| Enrichment culture persistent carriage | **0.007** | | |
| Positive | 3.87 | 38.46 | 42.58 |
| Negative | 3.47 | 61.54 | 57.42 |
| Hormonal Contraceptives (Women only, n = 505)[e] | Average Popularity (3.81) | Relative physical isolation (%) | Relative frequency all (%) |
| | **0.006** | | |
| Non-user | 4.00 | 58.97 | 64.88 |
| Progestin only | 2.65 | 13.16 | 3.97 |
| Low Estrogen | 3.44 | 13.16 | 9.92 |
| High Estrogen | 3.63 | 15.79 | 19.64 |

P-values are given next to variable names and represent a significant difference from the popularity average. P-values are calculated from t-test for two categories or ANOVA for more than two categories.

[a] Average popularity = Average number of friends nominating a participant as their friend

[b] Relative physical isolation = Number of participants not being nominated at all

[c] BMI = body mass index

[d] Physical activity: None = reading, watching TV, or other sedentary activity; Low level = walking, cycling, or other forms of exercise at least 4 hours a week; Medium level = participation in recreational sports, heavy outdoor activities with minimum duration of 4 hours a week; High level = Participation in heavy training or sports competitions regularly several times a week.

[e] Hormonal contraceptives: Non-user = No current use of hormonal contraceptives (women only); Progestin-only = Use of hormonal contraceptives with progestin (Cerazette, Nexplanon, Depo-provera, Implanon); Combination contraceptives low estradiol = Use of hormonal contraceptives with progestin and ethinyl estradiol less than or equal to 20µg (Mercilon, Yasminelle, Loette 28, Nuvaring). Combination contraceptives high estradiol = Use of hormonal contraceptives with progestin and ethinyl estradiol greater than or equal to 30µg (Marvelon, Yasmin, Microgynon, Oralcon, Diane, Synfase, Evra, Zyrona). Women taking contraceptives, but who were unable to recognize the brand were removed from the analysis

**Supplementary Tabl 6e** Average number of positive friends with respect to *Staphylococcus aureus* persistent nasal carrier status. The Fit Futures 1 study (N = 1038).

|  | Average number of friends | P-value[a] |
|---|---|---|
| **Direct culture** |  | **0.002** |
| Persistent carrier | 1.85 |  |
| Non-carrier | 1.56 |  |
| **Enrichment culture** |  | **< 0.001** |
| Persistent carrier | 2.54 |  |
| Non-carrier | 2.21 |  |
| [a]Student's t-test | | |

**Supplementary Table 7** Logistic regression model of *Staphylococcus aureus* persistent nasal carrier status with respect to positive friends. The Fit Futures 1 study (N = 1038).

|  | Estimate | Std Error | P-value |
|---|---|---|---|
| **Direct culture** |  |  |  |
| Intercept | - 1.09 | 0.11 | **<0.001** |
| Number of friends that are persistent carriers | 0.16 | 0.05 | **0.0016** |
| **Enrichment culture** |  |  |  |
| Intercept | - 0.63 | 0.12 | **<0.001** |
| Number of friends that are persistent carriers | 0.14 | 0.04 | **<0.001** |

**Supplementary Table 8  Attendance dates for each high school.** The Fit Futures 1 study (N = 1038).

| Week | Year | H1 | H2 | H3 | H4 | H5 | H6 | H7 | H8 | Friends |
|---|---|---|---|---|---|---|---|---|---|---|
| 38 | 2010 | 32 | 0 | 0 | 0 | 0 | 0 | 0 | 0 | 64.79 % |
| 39 | 2010 | 24 | 0 | 0 | 0 | 0 | 0 | 0 | 0 | 56.39 % |
| 40 | 2010 | 36 | 0 | 0 | 0 | 0 | 0 | 0 | 0 | 47.82 % |
| 41 | 2010 | 36 | 0 | 0 | 0 | 0 | 0 | 0 | 0 | 57.22 % |
| 42 | 2010 | 35 | 0 | 0 | 0 | 0 | 0 | 0 | 0 | 49.29 % |
| 43 | 2010 | 30 | 0 | 0 | 0 | 0 | 0 | 0 | 0 | 54.56 % |
| 44 | 2010 | 6 | 16 | 0 | 0 | 0 | 0 | 0 | 0 | 42.80 % |
| 45 | 2010 | 0 | 40 | 0 | 0 | 0 | 0 | 0 | 0 | 57.79 % |
| 46 | 2010 | 0 | 42 | 0 | 0 | 0 | 0 | 0 | 0 | 62.78 % |
| 47 | 2010 | 0 | 32 | 0 | 0 | 0 | 0 | 0 | 0 | 60.57 % |
| 48 | 2010 | 0 | 6 | 0 | 0 | 0 | 0 | 0 | 28 | 60.69 % |
| 49 | 2010 | 4 | 0 | 0 | 0 | 0 | 0 | 0 | 34 | 48.51 % |
| 50 | 2010 | 4 | 4 | 0 | 0 | 0 | 0 | 0 | 31 | 39.06 % |
| 51 | 2010 | 0 | 0 | 0 | 0 | 0 | 0 | 0 | 0 | 100.00 % |
| 52 | 2010 | 0 | 0 | 0 | 0 | 0 | 0 | 0 | 0 | 100.00 % |
| 1 | 2011 | 0 | 2 | 0 | 0 | 0 | 0 | 6 | 27 | 32.14 % |
| 2 | 2011 | 0 | 0 | 0 | 0 | 0 | 0 | 43 | 0 | 61.51 % |
| 3 | 2011 | 0 | 0 | 0 | 0 | 0 | 0 | 45 | 0 | 47.00 % |
| 4 | 2011 | 0 | 0 | 0 | 0 | 0 | 0 | 40 | 0 | 47.87 % |
| 5 | 2011 | 0 | 0 | 0 | 0 | 0 | 0 | 46 | 0 | 54.24 % |
| 6 | 2011 | 0 | 0 | 30 | 0 | 0 | 0 | 10 | 0 | 45.17 % |
| 7 | 2011 | 0 | 0 | 41 | 0 | 0 | 0 | 0 | 0 | 50.57 % |
| 8 | 2011 | 0 | 0 | 44 | 0 | 0 | 0 | 2 | 0 | 56.52 % |
| 9 | 2011 | 0 | 0 | 43 | 0 | 0 | 0 | 0 | 0 | 53.10 % |
| 10 | 2011 | 0 | 0 | 0 | 0 | 0 | 0 | 0 | 0 | 100 % |
| 11 | 2011 | 0 | 0 | 8 | 12 | 17 | 0 | 0 | 0 | 69.10 % |
| 12 | 2011 | 0 | 0 | 0 | 4 | 18 | 19 | 0 | 0 | 54.51 % |
| 13 | 2011 | 0 | 0 | 0 | 15 | 24 | 5 | 0 | 0 | 45.04 % |
| 14 | 2011 | 0 | 0 | 0 | 22 | 26 | 0 | 0 | 0 | 43.44 % |
| 15 | 2011 | 0 | 0 | 2 | 31 | 0 | 2 | 0 | 0 | 43.76 % |
| 16 | 2011 | 0 | 0 | 0 | 0 | 0 | 0 | 0 | 0 | 100.00 % |
| 17 | 2011 | 0 | 0 | 0 | 14 | 0 | 0 | 0 | 0 | 33.10 % |

The first attendance date was 2010-September-20[th], which corresponds to Week 38 of 2010. The last attendance date was 2011-April-27[th], which corresponds to week 17 of 2011. Notice the public holidays in Norway, during weeks 51 and 52 of 2010 is Christmas holidays, and week 16 of 2011 is Easter holiday. H1 to H8 correspond to each of the high school identifiers. The "Friends" column shows the average proportion of friends nominated by each participant who attended the Fit Futures 1 study in the same week as the subject himself/ herself. The weighted average for all weeks is 52.07%.